\documentclass[preprint,aps,tightenlines,showpacs]{revtex4}
\usepackage{bm}

\usepackage{graphicx,epsfig}

\usepackage{dcolumn}     
\newcommand{\bna}{{\bm\nabla}}
\newcommand{\bfr}{{\bm r}}

\newcommand{\bfk}{{\bm k}}
\newcommand{\bfq}{{\bm q}}
\newcommand{\bfp}{{\bm p}}

\newcommand{\bfj}{{\bm j}}
\newcommand{\bfL}{{\bm L}}
\newcommand{\bfs}{{\bm s}}
\newcommand{\bfS}{{\bm S}}

\def\beq{\begin{equation}}
\def\eeq{\end{equation}}
\def\beqy{\begin{eqnarray}}
\def\eeqy{\end{eqnarray}}

\begin{document}

{

\title{Quantum Monte Carlo calculations of magnetic moments and $\bm{M}$1 transitions
in $\bm{A\leq7}$ nuclei including meson-exchange currents}

\author{L. E. Marcucci$^{1,2}$}
\email{marcucci@df.unipi.it}
\author{Muslema Pervin$^3$}
\email{muslema@phy.anl.gov}
\author{\mbox{Steven C. Pieper$^3$}}
\email{spieper@anl.gov}
\author{R. Schiavilla$^{4,5}$}
\email{schiavil@jlab.org}
\author{R. B. Wiringa$^3$}
\email{wiringa@anl.gov}
\affiliation{
$^1$\mbox{Department of Physics ``Enrico Fermi", University of Pisa, I-56127 
Pisa, Italy}\\
$^2$INFN, Sezione di Pisa, I-56127 Pisa, Italy\\
$^3$Physics Division, Argonne National Laboratory, Argonne, Illinois 60439\\
$^4$Theory Center, Jefferson Laboratory, Newport News, Virginia 23606\\
$^5$Department of Physics, Old Dominion University, Norfolk, Virginia 23529
}

\date{\today}

\begin{abstract}
Green's function Monte Carlo calculations of magnetic moments and $M1$
transitions including two-body meson-exchange current (MEC) contributions
are reported for $A\leq7$ nuclei.
The realistic Argonne $v_{18}$ two-nucleon and Illinois-2 
three-nucleon potentials are used to generate the nuclear wave functions.
The two-body meson-exchange operators are constructed to satisfy the
continuity equation with the Argonne $v_{18}$ potential.
The MEC contributions increase the $A$=3,7 isovector magnetic moments by
16\% and the $A$=6,7 $M1$ transition rates by 17--34\%, bringing
them into very good agreement with the experimental data.
\end{abstract}

\pacs{21.10.-k, 23.20.-g, 23.40.-s}

\maketitle

}

\section {Introduction}
In a recent paper \cite{PPW07} we reported quantum Monte Carlo (QMC)
calculations of electroweak transitions in $A=6,7$ nuclei.
The QMC method is a two-step process, with an initial variational Monte 
Carlo (VMC) calculation to find a good trial function, followed by a 
Green's function Monte Carlo (GFMC) calculation to refine the solution.
When used with the Argonne $v_{18}$ two-nucleon~\cite{WSS95} ($N\!N$)
and Illinois-2 three-nucleon~\cite{PPWC01} ($3N$) potentials, the final GFMC results
reproduce the ground- and excited-state energies for $A\le10$ nuclei 
\cite{PW01,PVW02,PWC04,P05} very well.

In Ref.~\cite{PPW07} we studied magnetic dipole ($M1$) and electric
quadrupole ($E2$) transitions and nuclear beta-decay (Fermi and 
Gamow-Teller) rates.  These were the first off-diagonal
matrix element calculations using the nuclear GFMC method.  However, only
one-body transition operators were used to calculate the matrix elements.
We noted that two-body meson-exchange-current (MEC) operators are known
to increase isovector magnetic moments by 15-20\% for $A$=3 nuclei~\cite{Car98},
while a previous VMC calculation for the width of the first $M1$ transition
in $^6$Li was also increased by 20\%~\cite{WS98}.
In this paper we use GFMC wave functions to investigate MEC contributions 
to magnetic moments 
for the ground states of $A$=2--7 nuclei as well as a number of $M1$ 
transitions in $A$=6,7 nuclei.  We find significant isovector
contributions from the MEC operators and overall very good agreement
with experiment.

A brief review of the QMC calculational method is given in 
Sec.~\ref{sec:qmc}.
The electromagnetic current operator is discussed in detail in 
Sec.~\ref{sec:emc}.
Results and conclusions are given in Secs.~\ref{sec:res} and~\ref{sec:conclusions}.

\section {Quantum Monte Carlo method for transitions}
\label{sec:qmc}

We evaluate the diagonal magnetic moment matrix element $\langle
\Psi(J^{\pi};T)|{\cal O}| \Psi(J^\pi;T)\rangle$ and the off-diagonal
transition matrix element $\langle
\Psi_f(J^{\pi'};T')|{\cal O}| \Psi_i(J^\pi;T)\rangle$, where $\cal O$ is
the full electromagnetic operator.  The nuclear wave function with a
specific spin-parity $J^\pi$ and isospin $T$ is denoted as 
$\Psi(J^\pi;T)$ and is a solution of the many-body 
Schr\"{o}dinger equation
\beqy H \Psi(J^\pi;T)= E \Psi(J^\pi;T) \ .\eeqy
The Hamiltonian used here has the form
\beqy H = \sum_{i} K_i + {\sum_{i<j}} v_{ij} + \sum_{i<j<k}
V_{ijk} \ ,
\eeqy 
where $K_i$ is the nonrelativistic kinetic energy and $v_{ij}$ and $V_{ijk}$
are respectively the Argonne $v_{18}$ (AV18) \cite{WSS95} and Illinois-2
(IL2) \cite{PPWC01} potentials.
The VMC trial function $\Psi_T(J^\pi;T)$ for a given nucleus is constructed
from products of two- and three-body correlation operators acting on an
antisymmetric single-particle state of the appropriate quantum numbers.
The correlation operators are designed to reflect the influence of the
interactions at short distances, while appropriate boundary conditions
are imposed at long range~\cite{W91,PPCPW97}.
The $\Psi_T(J^\pi;T)$ has embedded variational parameters
that are adjusted to minimize the expectation value
\begin{equation}
 E_V = \frac{\langle \Psi_T | H | \Psi_T \rangle}
            {\langle \Psi_T   |   \Psi_T \rangle} \geq E_0 \ ,
\label{eq:expect}
\end{equation}
which is evaluated by Metropolis Monte Carlo integration~\cite{MR2T2}.
Here $E_0$ is the exact lowest eigenvalue of $H$ for the specified quantum numbers.
A good variational trial function has the form
\begin{equation}
   |\Psi_T\rangle = \left[1 + \sum_{i<j<k}\widetilde{U}^{TNI}_{ijk} \right]
                    \left[ {\cal S}\prod_{i<j}(1+U_{ij}) \right]
                    |\Psi_J\rangle \ .
\label{eq:psit}
\end{equation}
The Jastrow wave function, $\Psi_J$, is fully antisymmetric and has the
$(J^\pi;T)$ quantum numbers of the state of interest, while $U_{ij}$
and $\tilde{U}^{TNI}_{ijk}$ are the two- and three-body correlation
operators.  More details may be found in Ref.~\cite{PPW07}.
The error in the variational energy $E_{V}$ is of order
$|\Psi_{0}-\Psi_{T}|^{2}/|\Psi_{T}|^{2}$, 
where $\Psi_{0}$ is the exact lowest-energy eigenstate of $H$ for a given set 
of quantum numbers.
Other expectation values calculated with
$\Psi_{T}$ have errors of order $|\Psi_{0}-\Psi_{T}|/|\Psi_{T}|$.

The GFMC method~\cite{C87,C88} reduces the VMC errors by using the relation
\begin{equation}
\Psi_0 = \lim_{\tau \rightarrow \infty} \exp [ - ( H - E_0) \tau ] \Psi_T \ ;
\end{equation}
that is the operator $\exp [ - ( H - E_0) \tau ]$ 
projects $\Psi_{0}$ out of $\Psi_T$.
If the maximum $\tau$ actually used is large enough,
the eigenvalue $E_{0}$ is calculated exactly while other expectation values
are generally calculated neglecting terms of order $|\Psi_{0}-\Psi_{T}|^{2}/|\Psi_{T}|^{2}$
and higher~\cite{PPCPW97}.

In the following we present a brief overview of the nuclear GFMC method;
much more detail may be found in Refs.~\cite{PPCPW97,WPCP00}.
We start with the $\Psi_{T}$ of Eq.~(\ref{eq:psit}) and define the propagated
wave function $\Psi(\tau)$
\begin{eqnarray}
 \Psi(\tau) = e^{-({H}-E_{0})\tau} \Psi_{T}
            = \left[e^{-({H}-E_{0})\triangle\tau}\right]^{n} \Psi_{T} \ ,
\end{eqnarray}
where we have introduced a small time step, $\tau=n\triangle\tau$;
obviously $\Psi(\tau=0) =  \Psi_{T}$ and
$\Psi(\tau \rightarrow \infty) = \Psi_{0}$.
Quantities of interest are evaluated in terms of a ``mixed'' expectation value
between $\Psi_T$ and $\Psi(\tau)$:
\begin{eqnarray}
\langle O(\tau) \rangle_M & = & \frac{\langle \Psi(\tau) | O |\Psi_{T}
\rangle}{\langle \Psi(\tau) | \Psi_{T}\rangle}.
\label{eq:expectation}
\end{eqnarray}
The desired expectation values would, of course, have $\Psi(\tau)$ on both
sides; by writing $\Psi(\tau) = \Psi_{T} + \delta\Psi(\tau)$  and neglecting
terms of order $[\delta\Psi(\tau)]^2$, we obtain the approximate expression
\begin{eqnarray}
\langle O (\tau)\rangle =
\frac{\langle\Psi(\tau)| O |\Psi(\tau)\rangle}
{\langle\Psi(\tau)|\Psi(\tau)\rangle}
\approx \langle O (\tau)\rangle_M
    + [\langle O (\tau)\rangle_M - \langle O \rangle_V] ~,
\label{eq:pc_gfmc}
\end{eqnarray}
where $\langle O \rangle_{\rm V}$ is the variational expectation value.

For off-diagonal matrix elements relevant to this work the 
generalized mixed estimate is given by the expression
\begin{eqnarray}
 \frac{\langle\Psi^f(\tau)| O |\Psi^i(\tau)\rangle}{\sqrt{\langle \Psi^f(\tau) | \Psi^f(\tau)\rangle}
\sqrt{\langle \Psi^i(\tau) |\Psi^i(\tau)\rangle}}
\approx 
  \langle O(\tau) \rangle_{M_i}
+ \langle O(\tau) \rangle_{M_f}-\langle O \rangle_V \ ,
\label{eq:extrap}
\end{eqnarray}
where
\begin{eqnarray}
\langle O(\tau) \rangle_{M_i} 
& = & \frac{\langle \Psi^f_T | O |\Psi^i(\tau)\rangle}
           {\langle \Psi^i_T|\Psi^i(\tau)\rangle}
      \sqrt{\frac{\langle \Psi^i_T|\Psi^i_T\rangle}
           {\langle \Psi^f_T | \Psi^f_{T}\rangle}} \ , 
\label{eq:mixed_i}  \\
\end{eqnarray}
and $\langle O(\tau) \rangle_{M_f}$ is defined similarly.
For more details see Eqs.~(19-24) and the accompanying discussions in Ref.~\cite{PPW07}.

\section{The electromagnetic current operator}
\label{sec:emc}

The model used 
for the nuclear electromagnetic current operator
${\bfj}({\bfq})$ is based on the study of 
Ref.~\cite{Mar05}. It represents ${\bfj}({\bfq})$
as a sum of one-, two- and 
three-body terms that operate on the nucleon degrees of freedom, 
\begin{equation}
  \bfj(\bfq)=\sum_i \bfj_i(\bfq)+\sum_{i<j} \bfj_{ij}(\bfq)
            +\sum_{i<j<k} \bfj_{ijk}(\bfq)
   \ , \label{eq:current}
\end{equation}
${\bfq}$ being the three-momentum transfer.
The one-body operator ${\bfj}_i({\bfq})$ 
is derived from the nonrelativistic reduction
of the covariant single-nucleon current, by expanding
in inverse powers of the nucleon mass $m$. 
In the notation of Ref.~\cite{Car98}, it is
written as
\begin{equation}
  \bfj_i(\bfq)=\frac{\epsilon_i}{2m}\{ \bfp_i,{\rm e}^{{\rm i}\bfq\cdot\bfr_i}\}
               +\frac{\rm i}{2m}\mu_i{\bm \sigma}_i\times\bfq \,
                 {\rm e}^{{\rm i}\bfq\cdot\bfr_i}
  \ , \label{eq:cr1}
\end{equation}
where $\{\cdots , \cdots \}$ denotes the anticommutator, the quantities 
$\epsilon_i$ and $\mu_i$ are defined as
\begin{eqnarray}
  \epsilon_i&=&\frac{1}{2}\left[G_E^S(q_\mu^2) 
 +G_E^V(q_\mu^2)\tau_{i,z}\right]\ , \label{eq:ei} \\
  \mu_i&=&\frac{1}{2}\left[G_M^S(q_\mu^2) 
 +G_M^V(q_\mu^2)\tau_{i,z}\right]\ ,\label{eq:mi}
\end{eqnarray}
and $\bfp$, ${\bm \sigma}$ and ${\bm \tau}$ are the nucleon's 
momentum, Pauli spin and isospin operators, respectively. 
Finally, 
$G_E^S(q_\mu^2)$ ($G_M^S(q_\mu^2)$) and $G_E^V(q_\mu^2)$ ($G_M^V(q_\mu^2)$)
are the isoscalar and isovector combinations of the nucleon
electric (magnetic) Sachs form factors, respectively, evaluated at the
four-momentum transfer $q_\mu^2=q^2-\omega^2$ with $\omega=\sqrt{q^2+M_f^2}-M_i$,
where $M_i$ and $M_f$ are initial and final nuclear masses (only elastic 
scattering or inelastic scattering to discrete final states are considered
in the present work). 

The current operator satisfies the current conservation relation (CCR)
\begin{equation}
  {\bfq}\cdot{\bfj}({\bfq})= [H,\rho({\bfq})]\ .\label{eq:ccr}
\end{equation}
Here $H$ is the nuclear Hamiltonian consisting of two- and three-nucleon 
interactions, the AV18~\cite{WSS95} 
and IL2~\cite{PPWC01} potentials, respectively, and $\rho(\bfq)$ is the charge 
operator which, to lowest order in $1/m$, is written as
\begin{equation}
\rho(\bfq)=\sum_i\rho_i(\bfq) \ , \label{eq:ch}
\end{equation}
with
\begin{equation}
\rho_i(\bfq)=\epsilon_i\,{\rm e}^{{\rm i}\bfq\cdot\bfr_i} \ .
\end{equation}
To this order, the CCR separates into
\begin{eqnarray}
  {\bfq}\cdot{\bfj}_i({\bfq})&=& 
\biggl[{{\bfp}_i^2\over 2m},\rho_i({\bfq})\biggr]
    \ ,\label{eq:ccr1}\\
  {\bfq}\cdot{\bfj}_{ij}({\bfq})&=& [v_{ij},\rho_i({\bfq})+\rho_j({\bfq})]
 \ , \label{eq:ccr2}
\end{eqnarray}
and similarly for the three-body current ${\bfj}_{ijk}({\bfq})$.
The one-body current of Eq.~(\ref{eq:cr1})
is easily seen to satisfy Eq.~(\ref{eq:ccr1}). 

\subsection{Two-nucleon current}

The two-body current operator ${\bfj}_{ij}({\bfq})$ is
separated into two parts, labeled model-independent (MI)
and model-dependent (MD), following the scheme of Ref.~\cite{Ris89}. 
The MI two-body currents have longitudinal
components that satisfy the CCR of Eq.~(\ref{eq:ccr2}) 
with the $N\!N$ potential $v_{ij}$, {\it i.e.}~the 
AV18~\cite{WSS95}. The potential can be written as 
\begin{equation}
  v_{ij}=v^{IC}_{ij}+v^{IB}_{ij}\ , \qquad
  v^{IC}_{ij}=v^{0}_{ij}+v^{p}_{ij}\ ,
  \label{eq:vij}
\end{equation}
where $v_{ij}^{IC}$ and $v_{ij}^{IB}$ are the 
isospin-symmetry conserving ($I$$C$) and breaking ($I$$B$) 
parts of the 
potential, respectively, and 
$v_{ij}^0$ and $v_{ij}^p$ are the momentum-independent and 
momentum-dependent parts of the interaction. For the AV18, $v_{ij}^0$ 
corresponds to the contributions of the static components, including 
isospin-independent and isospin-dependent central, spin-spin, 
and tensor terms, while $v_{ij}^p$ 
retains the contributions from the spin-orbit and quadratic 
momentum-dependent components.  The $v_{ij}^{IB}$ part
in the AV18 is parameterized by the four operators 
\begin{equation}
  O^{p=15,\ldots,18}_{ij}=T_{ij}\>\>\> ,\,
  {\bm\sigma}_i\cdot{\bm\sigma}_j T_{ij}\>\>\>,\,
  S_{ij}\, T_{ij}\>\>\>, \, (\tau_{i,z}+\tau_{j,z}) \ ,
\label{eq:ocsb}
\end{equation}
where $S_{ij}$ is the standard tensor operator and the isotensor operator 
$T_{ij}$ is defined as $T_{ij}=3\, \tau_{i,z}\,\tau_{j,z}-{\bm\tau}_i\cdot
{\bm\tau}_j$.

The MI two-body currents arising from $v_{ij}^0$ 
have been constructed 
following the procedure of Ref.~\cite{Ris85}, 
which will be hereafter referred to as the meson-exchange (ME) scheme.
Within this scheme, the isospin-dependent static part of $v_{ij}^0$
is assumed to be induced by exchanges of effective pseudoscalar (PS), or 
``$\pi$-like,'' and vector (V), or``$\rho$-like,'' mesons.  The
propagators associated with these exchanges are projected out of the 
(isospin-dependent) central, spin-spin, and tensor components of $v_{ij}^0$.
The resulting two-body currents satisfy the 
CCR with $v_{ij}^0$ by construction. Explicit expressions can be found in a 
number of references (see Ref.~\cite{Mar05} and references therein).

The currents arising from $v_{ij}^p$ have been obtained following the 
procedure of Ref.~\cite{Sac48}, which will be referred to 
as the minimal-substitution (MS) scheme, reviewed and generalized
in Ref.~\cite{Mar05}.  We first note that 
the isospin operator ${\bm \tau}_i\cdot{\bm \tau}_j$ can be expressed
in terms of the space-exchange operator ($P_{ij}$), using the relation
\begin{equation}
  {\bm \tau}_i\cdot{\bm \tau}_j = -1-
  (1+{\bm \sigma}_i\cdot{\bm \sigma}_j) P_{ij} \ ,
  \label{eq:tt}
\end{equation}
valid when operating on antisymmetric wave functions.
The operator $P_{ij}$ is defined as 
$P_{ij}=   {\rm e}^{{\bfr}_{ji}\cdot{\bna}_i
  + {\bfr}_{ij}\cdot{\bna}_j}$, where 
the ${\bna}$-operators do not act on the 
vectors ${\bfr}_{ij}={\bfr}_i-{\bfr}_j=-{\bfr}_{ji}$ in the exponential.
In the presence of an electromagnetic
field, minimal substitution is performed both in the explicit
momentum dependence of the two-nucleon potential as well as
in the implicit momentum dependence implied by ${\bm \tau}_i\cdot{\bm \tau}_j$.  
The resulting current operators have been derived in Ref.~\cite{Mar05}. Here 
we only list the final result for the current operators 
associated with the isospin-independent and isospin-dependent 
spin-orbit interaction.  In this case, $v_{ij}^p$ can be expressed as
\begin{equation}
v_{ij}^p=v_{1,ij}^p+v_{2,ij}^p{\bm \tau}_i\cdot{\bm \tau}_j \ , 
\label{eq:vijp}
\end{equation}
where 
\begin{eqnarray}
  v^p_{1,ij}&=&v_{ls}(r)\, \bfL\cdot\bfS \ , \nonumber \\
  v^p_{2,ij}&=&v_{ls\tau}(r)\, \bfL\cdot\bfS\ ,
  \label{eq:so}
\end{eqnarray}
with $\bfL=\bfr_{ij}\times({\bfp}_i - {\bfp}_j )/2$, and $\bfS$ is 
the total spin of pair $ij$ and $v_{ls}$ and $v_{ls\tau}$ are the
spin-orbit parts of the $N\!N$ potential.  
Performing minimal substitution in $v^p_1$, we obtain 
\begin{equation}
  \bfj_{ij}(\bfq;ls)=
  \frac{1}{2} v_{ls}(r) \biggl( \epsilon_i\,{\rm e}^{{\rm i}\bfq\cdot\bfr_i}
  -\epsilon_j\, {\rm e}^{{\rm i}\bfq\cdot\bfr_j}\biggr)\,
  \bfS\times \bfr_{ij} \ .
  \label{eq:jsoq}
\end{equation}
For the isospin-dependent term $v^p_2$, we first symmetrize it as
\begin{equation}
v^p_{2,ij}\,{\bm\tau}_i\cdot{\bm\tau}_j=
\frac{1}{2}v_{ls\tau}(r) \left( \bfL\cdot\bfS
\, {\bm \tau}_i\cdot{\bm\tau}_j + {\bm \tau}_i\cdot{\bm\tau}_j\,
\bfL\cdot\bfS \right) \ ,
\label{eq:v2t}
\end{equation}
and the associated current then reads
\begin{eqnarray}
  \bfj_{ij}(\bfq;ls\tau)&=&\frac{1}{4} v_{ls\tau}(r)
  \bfS\times \bfr_{ij}\,
  \biggl( \eta_j {\rm e}^{{\rm i}\bfq\cdot\bfr_i} -
   \eta_i {\rm e}^{{\rm i}\bfq\cdot\bfr_j}\biggr) \nonumber \\
  &+&\frac{1}{2} v_{ls\tau}(r) G_E^V(q_\mu^2) 
  ({\bm\tau}_i\times{\bm \tau}_j)_z
  \biggl(\,\bfL\cdot\bfS
  \,\int_{\gamma_{ij}}d\bfs\,{\rm e}^{{\rm i}\bfq\cdot\bfs} 
  +\int_{\gamma^\prime_{ji}}d\bfs\,{\rm e}^{{\rm i}\bfq\cdot\bfs} \,\,
   \bfL\cdot\bfS\,\biggr) \ ,
  \label{eq:jsotq}
\end{eqnarray}
with $\eta_i=G_E^S(q_\mu^2)
\,{\bm\tau}_i\cdot{\bm\tau}_j+G_E^V(q_\mu^2)\tau_{i,z}$,
and $d{\bf s}$ is 
the infinitesimal step on the generic path $\gamma_{ij}$ 
($\gamma'_{ji}$) that 
goes from position $i$ ($j$) to position $j$ ($i$).
Since the choice of the two integration paths $\gamma_{ij}$ and 
$\gamma'_{ji}$ is arbitrary, 
the definition given above for $\bfj_{ij}(\bfq;ls\tau)$ is not unique. 
However, whatever choice is made, the corresponding current will 
satisfy the CCR with $v_{2,ij}^p$ by construction. 
The simplest choice for $\gamma_{ij}$ ($\gamma'_{ji}$) is that
of a linear path ($L$$P$), which leads to 
\begin{eqnarray}
  \bfj_{ij}^{LP}(\bfq;ls\tau)&=&\frac{1}{4} v_{ls\tau}(r)
  \bfS\times \bfr_{ij}\,
  \biggl( \eta_j {\rm e}^{{\rm i}\bfq\cdot\bfr_i} -
   \eta_i {\rm e}^{{\rm i}\bfq\cdot\bfr_j}\biggr) \nonumber \\
  &+&\frac{\rm i}{2} v_{ls\tau}(r) G_E^V(q_\mu^2) 
  ({\bm\tau}_i\times{\bm \tau}_j)_z
  \biggl[\,(\bfL\cdot\bfS)\,\, \bfr_{ij}\,f_{ij}(\bfq)
   +\bfr_{ij}\,f_{ij}(\bfq)\,
   (\bfL\cdot\bfS)\,\biggr] \ ,
\label{eq:jsotl}
\end{eqnarray}
where $f_{ij}(\bfq)$ is defined as
\begin{equation}
  f_{ij}(\bfq)=\frac{ {\rm e}^{ {\rm i}\bfq\cdot\bfr_i} 
        - {\rm e}^{ {\rm i}\bfq\cdot\bfr_j} }
  {\bfq\cdot\bfr_{ij} } \ ,
\label{eq:fijq}
\end{equation}
and $f_{ij}(\bfq=0)={\rm i}$.  It is interesting to
note that, in the limit ${\bfq}\rightarrow 0$, the two-body
current operator derived in the MS scheme becomes 
path-independent and hence unique~\cite{Mar05}. Therefore,
for processes involving small momentum transfers, the intrinsic
arbitrariness of the MS scheme may be of little consequence.

In earlier works, for example Refs.~\cite{Car90,Sch91,Viv96,Mar98,Viv00},
the two-body currents from the spin-orbit interaction 
were constructed within the ME scheme, by assuming
that its isospin-independent components are due to
exchanges of ``$\sigma$-like'' and ``$\omega$-like'' 
mesons, while the isospin-dependent ones originate from
``$\rho$-like'' exchanges.  The resulting currents, however,
are not exactly conserved.  This lack of consistency 
seems to lead to a significant discrepancy between theory
and experiment in some of the $pd$ radiative capture polarization
observables, specifically
the tensor polarization observables $T_{20}$ and $T_{21}$, 
at low energies~\cite{Mar05}. 

The MI two-body currents arising from the $I$$B$ terms
are generated by the operator ${\bm \tau}_i\cdot{\bm\tau}_j$ present 
in the isotensor operator $T_{ij}$, and are easily
constructed~\cite{Mar05}.  However, their contributions 
have been found to be negligibly small in the study of the
electromagnetic structure of $A$=2 and 3 nuclei.  This
is also the case in the present study of electromagnetic
transitions in $A$=6 and 7 nuclei.  

The MD part of the two-body current is purely transverse and therefore is
not constrained by the CCR.  The model adopted here includes the (isoscalar)
$\rho\pi\gamma$ and (isovector) $\omega\pi\gamma$ transition currents, as well as 
the currents due to excitation of intermediate $\Delta$ isobars. 
The latter are obtained within a nonperturbative treatment based
on the transition correlation operator approach, reviewed below
in Sec.~\ref{subsec:delta}.
\begin{table}[!hbt]
\caption{Values of the coupling constants
$f_{\pi N\!N}$, $g_{\rho N\!N}$, $g_{\omega N\!N}$,  
$g_{\rho\pi\gamma}$, and $g_{\omega\pi\gamma}$ 
and monopole form factor cutoffs $\Lambda_\pi$, $\Lambda_\rho$ and
$\Lambda_\omega$ used in the present work.}
\label{tb:MDconst}
\vspace{5mm}
\begin{ruledtabular}
\begin{tabular}{cccccccc}
$f_{\pi N\!N}^2/4\pi$ &  $g_{\rho N\!N}^2/4\pi$ & $g_{\omega N\!N}^2/4\pi$  & 
$g_{\rho\pi\gamma}$ & $g_{\omega\pi\gamma}$ & $\Lambda_\pi$ & 
$\Lambda_\rho$ & $\Lambda_\omega$ \\
\hline
0.075 & 0.55 & 16.96 & 0.56 & 0.63 & 0.75 GeV & 1.25 GeV & 1.25 GeV \\
\end{tabular}
\end{ruledtabular}
\end{table}

The $\rho\pi\gamma$ and $\omega\pi\gamma$ MD two-body currents 
are given by
\begin{eqnarray}
\bfj_{\rho\pi\gamma}(\bfk_i,\bfk_j)&=&
{\rm i}\frac{f_{\pi N\!N} g_{\rho N\!N} g_{\rho\pi\gamma}}{m_\pi m_\rho}
{\bm \tau}_i\cdot{\bm \tau}_j \, (\bfk_i\times\bfk_j)\nonumber \\
&\times&
 \biggl[
\frac{ {\bm \sigma}_i\cdot\bfk_i}{(k_i^2+m_\pi^2)(k_j^2+m_\rho^2)}-
\frac{ {\bm \sigma}_j\cdot\bfk_j}{(k_i^2+m_\rho^2)(k_j^2+m_\pi^2)}\biggr]
\ , \label{eq:jrpg} \\
\bfj_{\omega\pi\gamma}(\bfk_i,\bfk_j)&=&
{\rm i}\frac{f_{\pi N\!N} g_{\omega N\!N} g_{\omega\pi\gamma}}{m_\pi m_\omega}
\,(\bfk_i\times\bfk_j)\nonumber \\
&\times&
 \biggl[
\frac{ {\bm \sigma}_i\cdot\bfk_i}{(k_i^2+m_\pi^2)(k_j^2+m_\omega^2)}\tau_{i,z}-
\frac{ {\bm \sigma}_j\cdot\bfk_j}{(k_i^2+m_\omega^2)(k_j^2+m_\pi^2)}\tau_{j,z}
\biggr] \ . \label{eq:jopg}
\end{eqnarray}
Here $\bfk_i$ ($\bfk_j$) denotes the fractional momentum
transfer to nucleon $i$ ($j$), so that $\bfk_i+\bfk_j=\bfq$.
The $g_{\rho\pi\gamma}$, $g_{\rho N\!N}$, $g_{\omega\pi\gamma}$,
$g_{\omega N\!N}$, and $f_{\pi N\!N}$ are the $\rho\pi\gamma$,
$\rho N\!N$, $\omega\pi\gamma$, $\omega N\!N$, and $\pi N\!N$
coupling constants, while $m_\pi$, $m_\rho$
and $m_\omega$ are the pion, $\rho$- and $\omega$-meson masses, respectively.
Finally, monopole form factors at the pion and vector-meson 
vertices, given by
\begin{equation}
f_a(k)=\frac{\Lambda_a^2-m_a^2}{\Lambda_a^2+k^2} \, , \,\, a=\pi,\rho,\omega \ ,
\label{eq:monff}
\end{equation}
are introduced, to take into account the finite 
size of nucleons and mesons. 
The values of all the coupling constants and 
the cutoffs $\Lambda_a$ adopted in this work 
are listed in Table~\ref{tb:MDconst}. 
In particular, the $\rho\pi\gamma$ and $\omega\pi\gamma$
coupling constants are obtained from the measured
widths of the 
$\rho\rightarrow\pi+\gamma$~\cite{Ber80} and 
$\omega\rightarrow\pi+\gamma$~\cite{Che71} decays, while
the $\omega N\!N$ coupling constant and the cutoffs 
$\Lambda_\pi$, $\Lambda_\rho$ and $\Lambda_\omega$ are rather soft 
but still close to those inferred from models of the $N\!N$
potential.

In Ref.~\cite{Mar05}, the currents
induced by the three-nucleon interaction $V_{ijk}$
associated with $P$-wave two-pion exchange 
were also constructed.  However, their contribution
to $A$=3 observables was calculated to be quite small.  These currents
are neglected in the present study.

\subsection{Beyond Nucleons Only}
\label{subsec:delta}

The simplest description of the nucleus views it
as being made up of nucleons, and 
assumes that all other sub-nucleonic degrees of freedom may be eliminated 
in favor of effective many-body operators acting on the nucleons' 
coordinates. The validity of such a description is based on the success 
it has achieved in the quantitative prediction of many nuclear 
observables~\cite{Car98}.  However, it is interesting to consider 
corrections to this picture by including the degrees of freedom associated 
with nuclear resonances as additional constituents of the nucleus. 
When treating phenomena which do not involve explicitly meson production, 
it is reasonable to expect that the lowest excitation of the nucleon, 
the $\Delta$-isobar, plays a leading role.
In this approximation, the nuclear wave function is written as
\begin{equation}
\Psi_{N+\Delta} = \Psi(N\!N\cdots N) + \Psi^{(1)}(N\!N\cdots N\Delta) 
+ \Psi^{(2)}(N\!N\cdots N\Delta\Delta)+ \cdots \ ,
\label{eq:psiND}
\end{equation}
where $\Psi$ is the part of the total wave function consisting only of 
nucleons, $\Psi^{(1)}$ is the component in which a single nucleon 
has been converted into a $\Delta$-isobar, and so on.  The nuclear 
two-body interaction is taken as
\begin{equation}
v_{ij}=\sum_{B_{i},B_{j}=N,\Delta}\sum_{B_{i}^{'},B_{j}^{'}=N,\Delta} v_{ij}
(B_{i}B_{j}\rightarrow B_{i}^{'}B_{j}^{'}) \ ,
\label{eq:tbi}
\end{equation}
where transition interactions such as $v_{ij}(N\!N\rightarrow N\Delta)$, 
$v_{ij}(N\!N\rightarrow \Delta\Delta)$, etc. are responsible for generating 
$\Delta$-isobar admixtures in the wave function. The long-range part of 
$v_{ij}$ is due to pion-exchange, while its short- and intermediate-range 
parts, influenced by more complex dynamics, are constrained by fitting $N\!N$ 
scattering data at lab energy $\leq$ 400 MeV and deuteron 
properties~\cite{WSA84}.

Once the $N\!N$, $N\Delta$, and $\Delta\Delta$ interactions have been 
determined, the problem is reduced to solving the $N+\Delta$ 
coupled-channel Schr\"{o}dinger equation. However, this would 
involve a large number of $N+\Delta$ 
channels and therefore the practical implementation of this method is 
very difficult. In a somewhat simpler approach, 
known as the transition-correlation-operator (TCO) 
method~\cite{Sch92}, the nuclear wave function is written as
\begin{equation}
\Psi_{N+\Delta}=\left[{\cal{S}}\prod_{i<j}\left(1\,+\,U^{tr}_{ij}\right)
\right]\,\Psi \ ,
\label{eq:psitco}
\end{equation}
where $\Psi$ is the nucleons-only wave function, 
$\cal{S}$ is a symmetrizer, and the transition operators $U_{ij}^{tr}$ 
are defined as
\begin{equation}
U_{ij}^{tr}\,=\,U_{ij}^{N\Delta}\,+\,U_{ij}^{\Delta N}\,+\,U_{ij}^
{\Delta\Delta} \ , 
\label{eq:tcoop}
\end{equation}
\begin{eqnarray}
U_{ij}^{N\Delta}&=&\left[u^{\sigma\tau II}(r_{ij}){\bm\sigma}_{i}\cdot
{\bf S}_{j}\,+\,u^{t\tau II}(r_{ij})S_{ij}^{II}\right]\,{\bm\tau}_{i}
\cdot{\bf T}_{j} \ , \label{eq:tcops1}\\
U_{ij}^{\Delta\Delta}&=&\left[u^{\sigma\tau III}(r_{ij}){\bf S}_{i}\cdot
{\bf S}_{j}\,+\,u^{t\tau III}(r_{ij})S_{ij}^{III}\right]\,{\bf T}_{i}\cdot
{\bf T}_{j} \ .
\label{eq:tcops2}
\end{eqnarray}
Here, ${\bf S}_{i}$ and ${\bf T}_{i}$ are spin- and isospin-transition 
operators which convert nucleon $i$ into a $\Delta$-isobar, $ S_{ij}^{II}$ 
and $S_{ij}^{III}$ are tensor operators in which, respectively, the Pauli 
spin operators of either particle $i$ or $j$, and both particles $i$ and $j$ 
are replaced by corresponding spin-transition operators. The $U_{ij}^{tr}$ 
vanishes in the limit of large interparticle separations, since no 
$\Delta$ components can exist asymptotically.
The functions $u^{\sigma\tau II}(r)$, $u^{t\tau II}(r)$, etc.,
are obtained from two-body bound and low-energy 
scattering state solutions of the full $N+\Delta$ 
coupled channel problem, with the Argonne $v_{28}$ (AV28) model~\cite{WSA84}
as discussed in Ref.~\cite{Sch92}. 

We note that the perturbation theory (PT) 
description of $\Delta$-admixtures 
is equivalent to the replacements:
\begin{eqnarray}
U_{ij}^{N\Delta,{\rm PT}}&=&\frac{v_{ij}(N\!N\rightarrow N\Delta)}
{m-m_{\Delta}} \ , \label{eq:UNDpt} \\
U_{ij}^{\Delta\Delta,{\rm PT}}&=&\frac{v_{ij}(N\!N\rightarrow \Delta\Delta)}
{2(m-m_{\Delta})} \ , \label{eq:UDDpt} 
\end{eqnarray}
where the kinetic energy contributions in the denominators of 
Eqs.~(\ref{eq:UNDpt}) and~(\ref{eq:UDDpt}) have been neglected 
(static $\Delta$ approximation). Note that the transition interactions 
$v_{ij}(N\!N\rightarrow N\Delta)$ 
and $v_{ij}(N\!N\rightarrow \Delta\Delta)$ have the same operator structure 
as $U_{ij}^{N\Delta}$ and $U_{ij}^{\Delta\Delta}$ of Eqs.~(\ref{eq:tcops1}) 
and~(\ref{eq:tcops2}), but with 
the $u^{\sigma\tau\alpha}(r)$ and $u^{t\tau\alpha}(r)$ functions 
replaced by, respectively,
\begin{eqnarray}
v^{\sigma\tau\alpha}(r)&=&
\frac{(ff)_{\alpha}}{4\pi}\frac{m_{\pi}}{3}\frac{e^{-x}}{x}\,C(x) \ ,
\label{eq:uvst} \\
v^{t\tau\alpha}(r)&=&
\frac{(ff)_{\alpha}}{4\pi}\frac{m_{\pi}}{3}\left(1+\frac{3}{x}+\frac{3}{x^2}
\right)\frac{e^{-x}}{x}\,C^{2}(x) \ .
\label{eq:uvtt} 
\end{eqnarray}
Here $\alpha$ = II, III, $x\equiv m_{\pi}r$, $(ff)_{\alpha}=f_{\pi N\!N}
f_{\pi N \Delta}$, $f_{\pi N \Delta}f_{\pi N \Delta}$ for $\alpha$ = II, III, 
respectively, and the cutoff function $C(x)\,=\,1-e^{-\lambda x^{2}}$, 
with $\lambda=4.09$.  In the AV28
model~\cite{WSA84} $f_{\pi N \Delta}=2f_{\pi N\!N}$.
This perturbative treatment has been often 
used in the literature to estimate the effect of $\Delta$ degrees of freedom 
on electroweak observables. However, it may lead to a substantial 
over prediction of their importance~\cite{Viv96,Sch92}, 
since it produces $N\Delta$ and $\Delta\Delta$ wave functions 
which are too large at short distance.

The nuclear electromagnetic current is now expanded into a sum
of many-body terms as in Eq.~(\ref{eq:current}). However, here
each term operates not only on the nucleon, but also on the
$\Delta$-isobar degrees of freedom. Therefore, the one- and
two-body currents (ignoring three-body currents) are written as 
\begin{eqnarray}
{\bfj}^{(1)}_{i}({\bfq})&=&\sum_{B,B^{'}=N,\Delta}{\bfj}_{i}({\bfq};
B\rightarrow B^{'}) \ , \\
\label{eq:1bterm}
{\bfj}_{ij}^{(2)}({\bfq})&=&{\sum}_{B_{i}, B_{j}=N, \Delta}
{\sum}_{B_{i}^{'}, B_{j}^{'}=N, \Delta}\,\, {\bfj}_{ij}({\bfq}; B_{i}B_{j}
\rightarrow B_{i}^{'}B_{j}^{'}) \ . \label{2bterm}
\end{eqnarray}
In the present work, however, we only keep the purely
nucleonic two-body currents discussed in the previous section.

The one-body $N\rightarrow\Delta$ transition and $\Delta$ currents are 
given by
\begin{eqnarray}
{\bfj}_{i}({\bfq};N\rightarrow \Delta)&=&-\frac{\rm i}{2 m}
G_{\gamma N\Delta}(q_{\mu}^{2}) {\rm e}^{{\rm i}{\bfq}\cdot{\bfr}_{i}} 
{\bfq}\times{\bf S}_{i} T_{z,i} \ , \label{eq:j1bND} \\
{\bfj}_{i}({\bfq};\Delta\rightarrow \Delta)&=&-\frac{\rm i}{24 m}
G_{\gamma\Delta\Delta}(q_{\mu}^{2}) {\rm e}^{{\rm i}{\bfq}\cdot{\bfr}_{i}} 
{\bfq}\times{\bm\Sigma}_{i} (1+\Theta_{z,i}) \ , \label{eq:j1bDD}
\end{eqnarray}
where ${\bm\Sigma}$ (${\bm\Theta}$) is the Pauli operator for the 
$\Delta$ spin 3/2 (isospin 3/2), and the expression for ${\bfj}_{i}({\bfq};
\Delta\rightarrow N)$ is obtained from that for ${\bfj}_{i}({\bfq};
N\rightarrow \Delta)$ by replacing the transition spin and isospin 
operators by their hermitian conjugates. The $N\Delta$-transition and 
$\Delta$ electromagnetic form factors, respectively $G_{\gamma N\Delta}$ 
and $G_{\gamma\Delta\Delta}$, are parameterized as
\begin{eqnarray}
G_{\gamma N\Delta}(q_{\mu}^{2})&=&\frac{\mu_{\gamma N\Delta}}
{\left(1+q_{\mu}^{2}/\Lambda^{2}_{N\Delta,1}\right)^{2}\sqrt{1+q_{\mu}^{2}/
\Lambda^{2}_{N\Delta,2}}} \ , \label{eq:gff1}\\
G_{\gamma \Delta\Delta}(q_{\mu}^{2})&=&\frac{\mu_{\gamma \Delta\Delta}}
{\left(1+q_{\mu}^{2}/\Lambda^{2}_{\Delta\Delta}\right)^{2}} \ .  
\label{eq:gff2}
\end{eqnarray}
Here the $N\Delta$-transition magnetic moment $\mu_{\gamma N\Delta}$ is 
taken equal to 3 $\mu_{N}$, as obtained from an analysis of $\gamma N$ 
data in the $\Delta$-resonance region~\cite{Car86}; this analysis also 
gives $\Lambda_{N\Delta,1}\,=$ 0.84 GeV and $\Lambda_{N\Delta,2}\,=$ 1.2 GeV. 
The value used for the $\Delta$ magnetic moment $\mu_{\gamma\Delta\Delta}$ 
is 4.35 $\mu_{N}$ by averaging results of a soft-photon analysis of 
pion-proton bremsstrahlung data near the 
$\Delta^{++}$ resonance~\cite{Lin91}, 
and $\Lambda_{\Delta\Delta}\,=$ 0.84 GeV as in the dipole 
parameterization of the nucleon form factor. 
In principle, $N$ to $\Delta$ 
excitation can also occur via an electric quadrupole transition. 
Its contribution, however, has been ignored, since the associated pion 
photoproduction amplitude is found to be experimentally small at 
resonance~\cite{Eri88}. Also neglected is the $\Delta$ convection current. 

\subsection{Matrix elements}
\label{subsec:calc}

\begin{figure}[!hbt]
\caption[Figure]{\label{fig:1bD}Diagrammatic representation of operators
included in $j(\Delta)$ due to one-body currents $j^{(1)}
(N\rightarrow\Delta)$, $j^{(1)}(\Delta\rightarrow N)$ and $j^{(1)}
(\Delta\rightarrow\Delta)$, and transition correlations $U^{N\Delta}$, 
$U^{\Delta N}$, $U^{\Delta\Delta}$, and corresponding hermitian conjugates.
Wavy, thin, thick, dashed and dashed with a $\times$ lines denote photons, nucleons,
$\Delta$-isobars, and transition correlations $U^{B B^{'}}$ and
${U^{B B^{'}}}^{\dagger}$, respectively.}
\includegraphics[height=8cm]{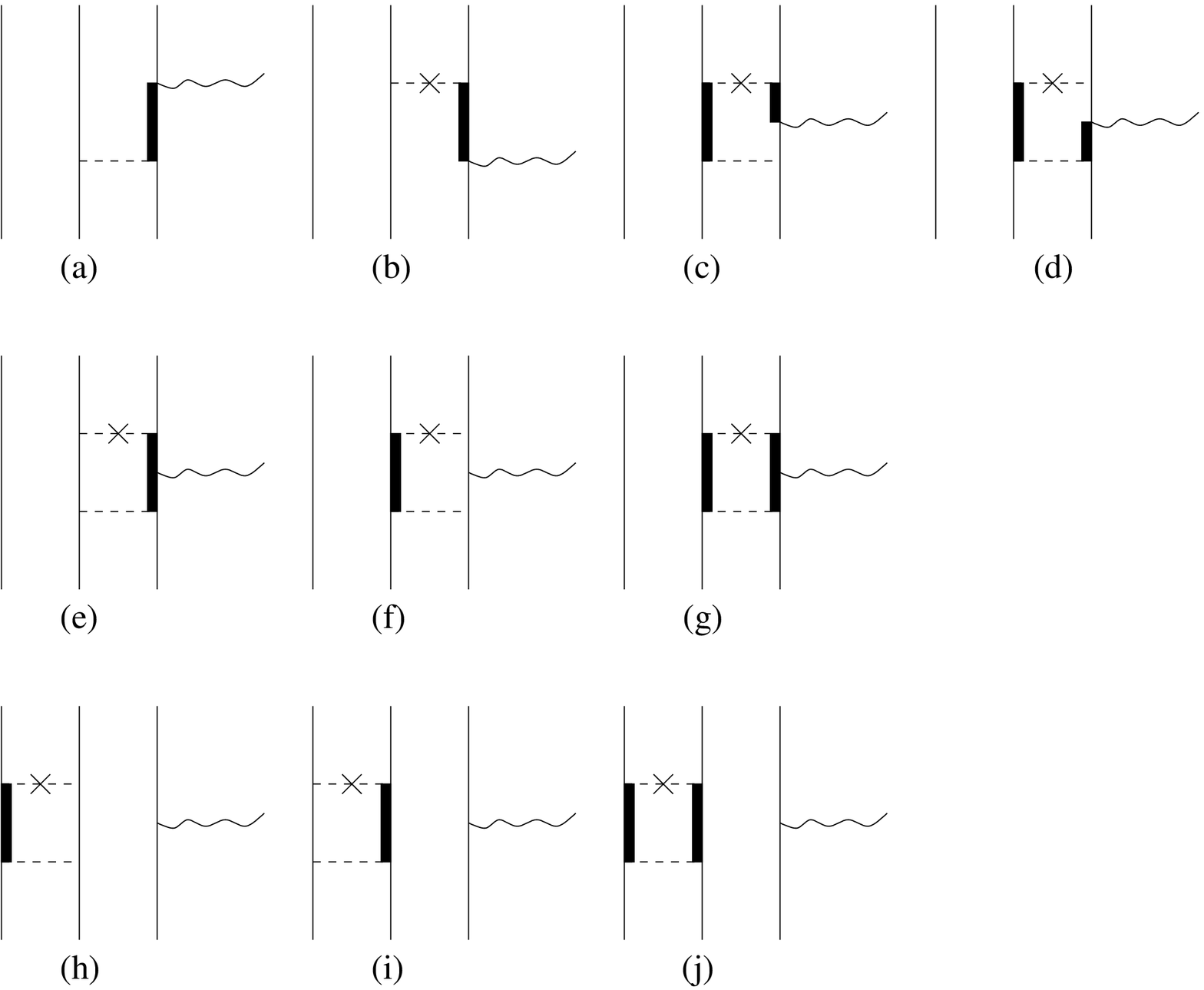}
\end{figure}

\begin{figure}[!hbt]
\caption[Figure]{\label{fig:3bDnegl} Diagrams associated with connected 
three-body terms, which are neglected in the present work. Wavy, thin,
thick, dotted, and dashed and dashed with a $\times$ lines denote photons, nucleons,
$\Delta$-isobars, the two-body current
$j^{(2)}(N\!N\rightarrow N\!N)$, and the transition correlations $U^{B B^{'}}$ and
${U^{B B^{'}}}^{\dagger}$ respectively.}
\includegraphics[height=3cm]{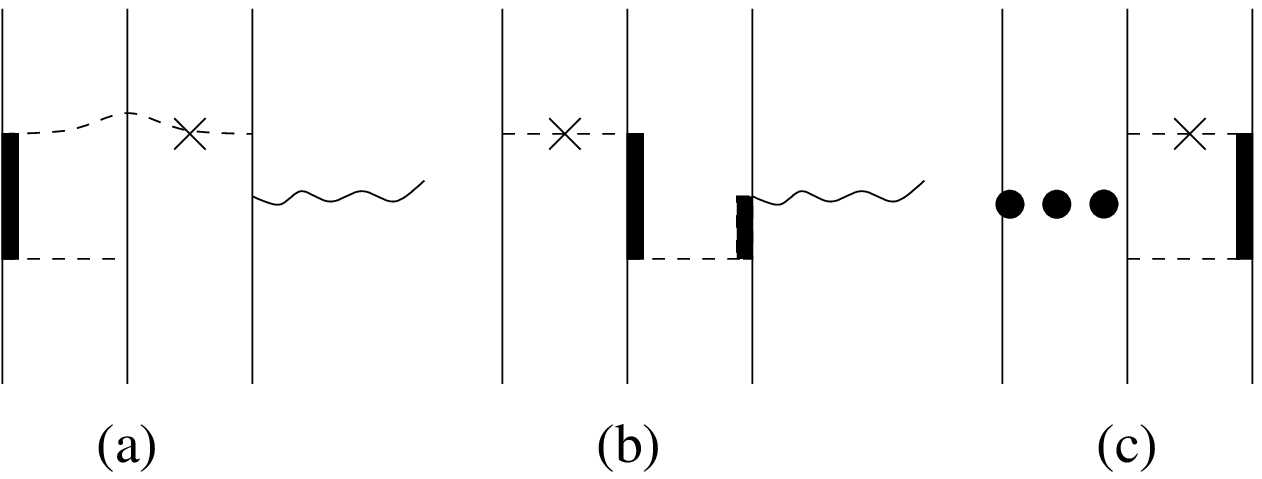}
\end{figure}

Matrix elements of the current operator can be written schematically as
\begin{equation}
j_{fi}=\frac{\langle\Psi_{N+\Delta,f}|j|\Psi_{N+\Delta,i}\rangle}
{[\langle\Psi_{N+\Delta,f}|\Psi_{N+\Delta,f}\rangle
\langle\Psi_{N+\Delta,i}|\Psi_{N+\Delta,i}\rangle]^{1/2}} \ ,
\label{eq:jfi}
\end{equation}
where the initial and final state wave functions 
include both nucleonic and $\Delta$-isobar 
degrees of freedom.  It is convenient to expand $\Psi_{N+\Delta}$ as 
\begin{equation}
\Psi_{N+\Delta} = \Psi + \sum_{i<j}U_{ij}^{tr}\Psi + 
\ldots \ ,
\label{eq:psind}
\end{equation}
and the matrix element of the current operator becomes
\begin{equation}
\langle\Psi_{N+\Delta,f}\,|\,j\,|\,\Psi_{N+\Delta,i}\rangle\,=\,\langle
\Psi_{f}\,|\,j(N\,{\rm only})\,|\,\Psi_{i}\rangle 
\,+\,\langle\Psi_{f}\,|\,j(\Delta)\,|\,\Psi_{i}\rangle \ .
\label{eq:schemj}
\end{equation}
Here $j(N \,{\rm only})$ denotes all one- and two-body contributions to 
${\bfj}({\bfq})$ which only involve nucleon degrees of freedom, i.e., 
$j(N \,{\rm only})\,=\,j^{(1)}(N\rightarrow N)\,+\,j^{(2)}(N\!N\rightarrow N\!N)$. 
The operator $j(\Delta)$ includes terms involving the $\Delta$-isobar 
degrees of freedom, associated with the explicit $\Delta$ currents $j^{(1)}
(N\rightarrow\Delta)$, $j^{(1)}(\Delta\rightarrow N)$ and $j^{(1)}(\Delta
\rightarrow\Delta)$, and with the 
transition operators $U_{ij}^{tr}$. The operator $j(\Delta)$ 
is illustrated 
diagrammatically in Fig.~\ref{fig:1bD}. The terms (a)-(g) 
in Fig.~\ref{fig:1bD} are two-body current 
operators, while the terms (h)-(j) are to be 
interpreted as renormalization corrections to the ``nucleonic'' matrix 
elements $\langle\Psi_{f}\,|\,j(N {\rm only})\,|\,\Psi_{i}\rangle$, due 
to the presence of $\Delta$-admixtures in the wave functions.
We note that not included in $j(\Delta)$ are all remaining connected 
three-body contributions of the type of Fig.~\ref{fig:3bDnegl}, 
which are expected to be significantly smaller than those considered 
in Fig.~\ref{fig:1bD}.

The terms in Fig.~\ref{fig:1bD} are expanded as operators acting on the 
nucleons' coordinates. For example, the terms (a) and (e) 
have the structure, respectively,
\begin{eqnarray}
({\rm a})&=&j_{i}^{(1)}(\Delta\rightarrow N)\,U_{ij}^{\Delta N} \ , 
\label{eq:aterm} \\
({\rm e})&=&{U_{ij}^{\Delta N}}^{\dagger}\,j^{(1)}_{i}
(\Delta\rightarrow\Delta)\,U_{ij}^{\Delta N} \ , \label{eq:eterm}
\end{eqnarray}
which can be reduced to operators involving only Pauli spin and isospin 
matrices by using the identities
\begin{eqnarray}
{\bf S}^{\dagger}\cdot{\bf A}\,{\bf S}\cdot{\bf B}&=&\frac{2}{3}{\bf A}
\cdot{\bf B}-\frac{\rm i}{3}{\bm\sigma}\cdot ({\bf A}\times{\bf B}) \ ,
\label{eq:SdagS} \\
{\bf S}^{\dagger}\cdot{\bf A}\,{\bm\Sigma}\cdot{\bf B}\,{\bf S}
\cdot{\bf C}&=&\frac{5}{3}\,{\rm i}\,{\bf A}\cdot({\bf B}\times{\bf C})-
\frac{1}{3}{\bm\sigma}\cdot{\bf A}\,{\bf B}\cdot{\bf C} \nonumber \\
& & -\frac{1}{3}{\bf A}\cdot {\bf B}\,{\bf C}\cdot{\bm\sigma}+\frac{4}{3}
{\bf A}\cdot({\bf B}\cdot{\bm\sigma}){\bf C} \ ,
\label{eq:SdagSigmaS}
\end{eqnarray}
where ${\bf A}$, ${\bf B}$ and ${\bf C}$ are vector operators that commute 
with ${\bm\sigma}$, but not necessarily among themselves. Expressions 
for the other terms of Fig.~\ref{fig:1bD} are obtained in a similar 
fashion.

The denominator of Eq.~(\ref{eq:jfi}) requires the calculation of the 
initial and final state wave function renormalizations, which 
are given by
\begin{eqnarray}
(N^\Delta)^2 =
\frac{\langle \Psi_{N+\Delta}\,|\,\Psi_{N+\Delta}\rangle}{\langle \Psi | \Psi \rangle} &=&
\frac{\langle \Psi\,|\,1\,+\,\sum_{i<j}[2\,{U_{ij}^{\Delta N}}^{\dagger}
U_{ij}^{\Delta N}\,+\,{U_{ij}^{\Delta \Delta}}^{\dagger}U_{ij}^{\Delta 
\Delta}]
\,|\,\Psi\rangle}{\langle \Psi | \Psi \rangle} \nonumber \\
&+&\,(\,{\rm three\!-\!body\, terms})
\label{eq:normal}
\end{eqnarray}
and the three-body terms have been neglected consistently with the 
approximation introduced in Eq.~(\ref{eq:schemj}), as discussed above.  
The wave-function renormalizations ($N^\Delta$)
for the different nuclei considered 
in the present work are listed in Table~\ref{tb:norm}.
The TCO approximation of Eq.~(\ref{eq:psind}) gives a renormalization of
1.0023 for the deuteron, whereas the exact coupled-channel result
for the AV28 potential is 1.0026.
Note that PT estimates of the importance of $\Delta$-isobar 
degrees of freedom in photo- and electro-nuclear observables typically 
include only the contribution from single $N\rightleftharpoons\Delta$ 
transitions (namely diagrams (a) and (b) in Fig.~\ref{fig:1bD}) and ignore 
the change in the wave function normalization. 

\begin{table}[tb]
\caption{Wave function renormalizations,  $N^\Delta$,
obtained for the $A$=2--7 nuclei, 
when the TCO calculation is based on the AV28 interaction is used with
purely nucleonic GFMC wave functions for the 
AV18+IL2 Hamiltonian model.}
\label{tb:norm}
\vspace{5mm}
\begin{ruledtabular}
\begin{tabular}{cccccc}
$^2$H  & $^3$H & $^3$He & $^6$Li & $^7$Li & $^7$Be \\
\colrule
1.0023 & 1.016 & 1.016  & 1.050  & 1.071 & 1.073(1)  \\
\end{tabular}
\end{ruledtabular}
\end{table}

\section {Results}
\label{sec:res}

In Ref.~\cite{PPW07} we reported results for fifteen electroweak transitions
in $A$=6,7 nuclei.  We pointed out that MEC contributions are
expected to be significant in magnetic moment ($\mu$) and
magnetic dipole ($M1$) transition calculations.
Here we present our results for $\mu$ for $A$=2--7 and $M1$ transitions for
$A$=6,7 including the MEC contributions. 
The first subsection discusses the magnetic moment results for $A$=2--7
nuclei and the second subsection discusses the $M1$ transitions in $A$=6,7 
nuclei.

The calculation of $\mu$'s or $M1$ transitions
is fairly straightforward.  For example, $\mu$ is obtained
from the diagonal matrix element
\begin{equation}
\mu  = -i \lim_{q\to 0} \frac{2\, m}{q}
\langle J^\pi,M_J\!=\!J;T\mid j_y(q\, \hat{\bf x}) \mid J^\pi,M_J\!=\!J;T\rangle \ ,
\end{equation}
where the momentum transfer ${\bf q}$ is taken along the $\hat{\bf x}$ axis,
the nuclear state with angular momentum quantized along the $\hat{\bf z}$ axis
has $M_J$=$J$, and $m$ is the nucleon mass.  The VMC or GFMC wave
function for the given $(J^\pi,M_J=J;T)$ state is then constructed.  Evaluation of
the various contributions is done as a function of the momentum transfer $q$
for several small values $q<0.05$ fm$^{-1}$ and then extrapolated smoothly
to the limit $q$=0.  The error due to extrapolation
is much smaller than the statistical error in the Monte Carlo sampling.

In the tables below, we present the one-body, i.e. impulse approximation (IA), results
and contributions from various pieces of the two-body MEC operators
separately: pseudoscalar + vector (PS+V), minimal-substitution (MS),
model-dependent (MD), and $\Delta$.  The $\Delta$ column includes contributions
from both the explicit MEC terms of Eqs.~(\ref{eq:j1bND}) and~(\ref{eq:j1bDD})
and the renormalization  of the nucleons-only terms as given by Eq.~(\ref{eq:jfi}):
\begin{equation}
\Delta=\frac{\langle\Psi_f|j(\Delta)+j(N \,\, {\rm only})|\Psi_i\rangle}
{N^\Delta_f\, N^\Delta_i\, \langle \Psi_f | \Psi_f \rangle\, \langle \Psi_i | \Psi_i \rangle }
- \frac{\langle\Psi_f|j(N \,\, {\rm only})|\Psi_i\rangle} 
{\langle \Psi_f | \Psi_f \rangle\, \langle \Psi_i | \Psi_i \rangle } \ .
\label{eq:rcc1}
\end{equation}

\subsection {Magnetic Moments in $A$=2--7 Nuclei}

Table~\ref{tb:mu} shows the magnetic moment results for $A$=2--7
nuclei.  The last two columns list the total magnetic moments and
corresponding experimental numbers~\cite{exp567,WN08}.
The total $\mu$ is obtained from the sum of the IA and the two-body
contributions from various pieces.  Apart from the $^2$H case, we
present the VMC results followed by the GFMC results in the following
row for each magnetic moment.  Hyperspherical harmonics (HH) results for
the trinucleons interacting by the AV18 $N\!N$ potential and older 
Urbana-IX (UIX)~\cite{PPCW95} $3N$ potential are shown for comparison.
We also present the GFMC isoscalar and isovector combinations for $A$=3,7.

\begin{table}[bt]
\caption{Magnetic moments in nuclear magnetons for $A=$2--7 nuclei; IA, PS, V, MS, MD
stand for impulse approximation, pseudoscalar, vector,
minimal-substitution, and model-dependent, respectively. 
Details can be found in the text.
IS and IV in the ``Nucleus'' column denote the isoscalar and
isovector combinations of the preceding nuclei.
The experimental values are from the compendia~\cite{exp567} except
for the very recent measurement for $^7$Be~\cite{WN08}; they have been
rounded to 3 decimal digits, except for $^2$H.}
\label{tb:mu}
\vspace{5mm}
\begin{ruledtabular}
\begin{tabular}{llccccccc}
Nucleus& Method & IA & \multicolumn{4}{c}{MEC}      & Total & Expt.\\
       &        &    &  PS$+$V & MS & MD & $\Delta$ &       & \\
\colrule
$^2$H &     & ~0.8467    & ~0         & -0.0022   & ~0.0031  &0.0009 & ~0.8485   & ~0.8574\\
\\
$^3$H &VMC  & ~2.580     & ~0.319     & -0.002    & ~0.017   &0.018  & ~2.932(1) & ~2.979   \\
$^3$H &GFMC & ~2.573(2)  & ~0.322(2)  & -0.002    & ~0.017   &0.014  & ~2.924(3) & ~2.979  \\ 
$^3$H &HH   & ~2.575     & ~0.321     & -0.001    & ~0.017   &0.014  & ~2.926    & ~2.979\\
$^3$He&VMC  & -1.766     & -0.317     & -0.001    & -0.010   &-0.013 & -2.107(1) & -2.128 \\
$^3$He&GFMC & -1.756(2)  & -0.318(2)  & -0.001    & -0.010   &-0.018 & -2.103(3) & -2.128 \\
$^3$He&HH   & -1.764     & -0.316     & -0.001    & -0.010   &-0.014 & -2.105    & -2.128 \\
IS &  GFMC  & 0.408	 & 0.001      & 0.002 	  & 0.003    &-0.003 & 0.411	 & 0.426\\
IV &  GFMC  & 4.329	 & 0.640      & 0.001     & 0.027    &0.030  & 5.027	 & 5.107\\
\\
$^6$Li&VMC  & ~0.815(1)  & ~0         & -0.008    & ~0.004   &-0.006 & ~0.805(1) & ~0.822 \\	  
$^6$Li&GFMC & ~0.810(1)  & ~0         & -0.007    & ~0.005   &-0.008 & ~0.800(1) & ~0.822 \\	  
\\
$^7$Li&VMC  & ~2.906(4)  & ~0.318(3)  & -0.011(1) & ~0.019   &-0.042 & ~3.190(7) & ~3.256 \\ 
$^7$Li&GFMC & ~2.870(8)  & ~0.340(6)  & -0.009(4) & ~0.020   &-0.053 & ~3.168(13)& ~3.256  \\
$^7$Be&VMC  & -1.098(5)  & -0.317(6)  & ~0.005    & -0.012   &-0.078 & -1.503(5) & -1.400\\
$^7$Be&GFMC & -1.058(9)  & -0.343(6)  & ~0.007    & -0.011   &-0.088 & -1.493(15) & -1.400\\
IS &  GFMC  & 0.906	 & 0.001      & -0.002 	  & 0.004    &-0.073 &  0.836	 & 0.929\\
IV &  GFMC  & 3.928	 & 0.683      & -0.019    & 0.030    &0.039  &  4.661	 & 4.654\\
\end{tabular}
\end{ruledtabular}
\end{table}

The results presented in Table~\ref{tb:mu} show
the significant impact of the two-body operators in those cases
with nuclear isospin $T=1/2$.  MEC contributions boost the IA by
about 16$\%$ in the $A$=3 isovector case and by about 19$\%$ in the $A$=7
nuclear states.  For the two $T=0$ states, namely $^2$H and $^6$Li,
we see that the IA magnetic moments are not modified
significantly by the MEC, as expected for any isoscalar state.
We note, however, that the present isoscalar MEC contributions
to the deuteron magnetic moment are smaller than reported
previously in Ref.~\cite{WSS95}.  This is the result of
the different way in which two-body currents
from the momentum dependent components of the AV18 have been
constructed in this work (see Ref.~\cite{Mar05}
for a discussion of this issue).

The magnetic moments for $^3$H and $^3$He are closer
to the experimental values of $\mu$ in both cases when the MEC
contributions are added to the IA values.  The major contributions
come from the pseudoscalar and vector piece of the currents.
The model-dependent piece and $\Delta$ contributions are small, but not
insignificant, while the minimal substitution piece is tiny.  
We also note that the VMC, GFMC, and HH results are all very close to
each other for all the separate pieces as well as for the total $\mu$.

The simplest picture of $^3$H consists of a $S=0$ pair of neutrons
and a proton, all in a total $L=0$ state.  That of $^3$He is the same with
proton and neutron interchanged.  In this picture the impulse approximation magnetic
moments of $^3$H and $^3$He would be the same as those of the proton
and neutron respectively: 2.79 and --1.91.  As can be seen in Table~\ref{tb:mu},
the GFMC impulse magnetic moments are 8\% smaller in magnitude than these.  
This is largely due to the tensor force which has several effects:
1) the wave functions contain $\sim 9\%$ of $L=2$ components which result in
orbital contributions of +0.04 and +0.05 to the magnetic moments
of $^3$H and $^3$He, respectively; 2) the odd nucleon is not 100\%
aligned with the nuclear spin; and 3) the pair of like nucleons has
$\sim 10\%$ $S=1$ component~\cite{FPPWSA96}.  The last two effects reduce the spin
contributions to the magnetic moments from the pure nucleon values
to 2.53 and --1.81.

A similar analysis of the $A$=7 magnetic moments can be made.
The ground state of $^7$Li looks a great deal
like an $\alpha$ particle plus triton in relative $P$-wave motion, 
whereas $^7$Be looks like $\alpha$
plus $^3$He.  Thus, the orbital contribution to the $A$=7 impulse
magnetic moments is expected to be significantly larger than for $A$=3; it is
0.42 and 0.67 for $^7$Li and $^7$Be, respectively. The
spin contributions, 2.43 and --1.72 are quenched relative
to the spin contributions to the $A$=3 values; this is probably
due to tensor interactions between a nucleon in the $\alpha$ core and
one in the valence $A$=3 cluster.  The MEC contributions are 
generally larger in the $A$=7 nuclei, again probably due to 
interactions between the core and valence nucleons.

\begin{figure}[ht!]
\centerline{\epsfig{file=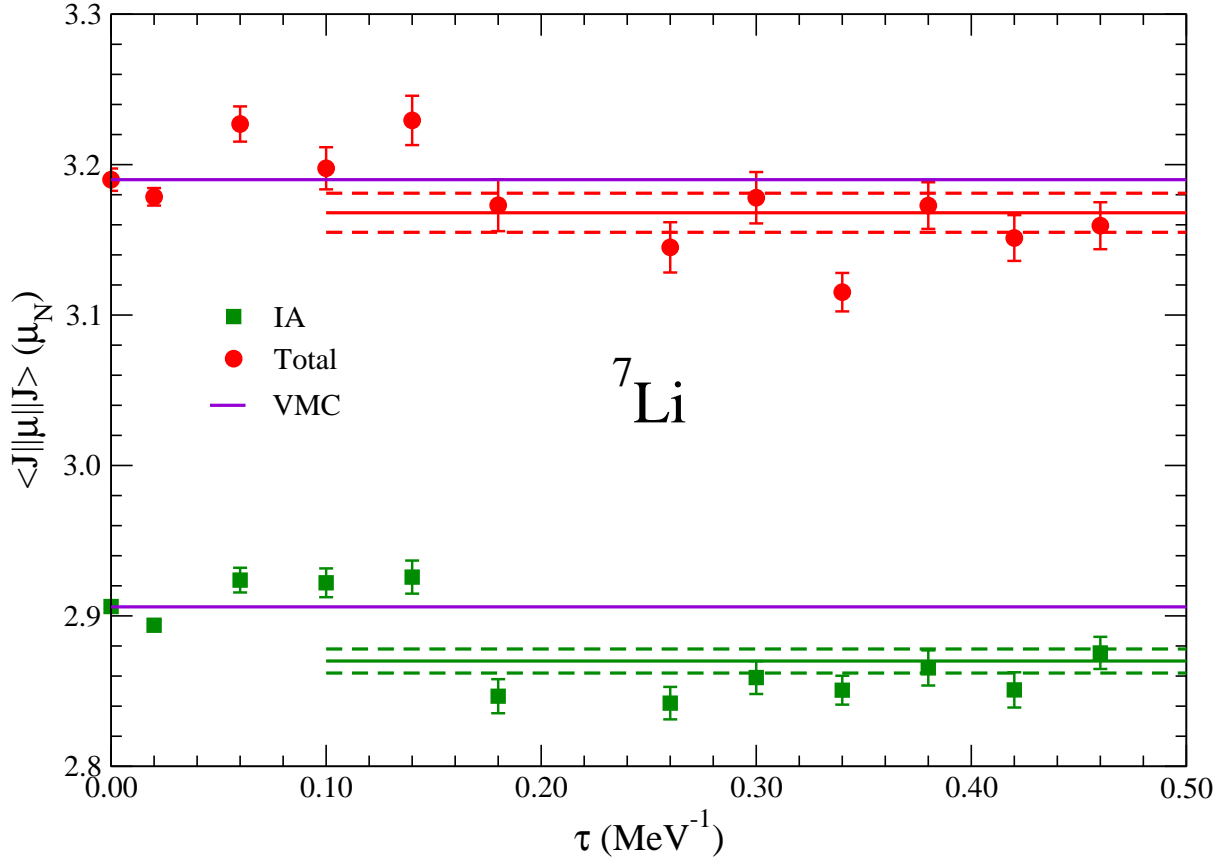,width=16.4cm}}
\caption{(Color online)
Extrapolated GFMC magnetic moment for the $^7$Li$(\frac{3}{2}^-)$ ground 
state in impulse approximation (green squares) and with MEC (red circles).
VMC values (purple lines) and averaged GFMC values (lines with error bars)
are also shown.}
\label{muli7}
\end{figure}

The GFMC propagation for the magnetic moment of the $^7$Li$(\frac{3}{2}^-)$
ground state is shown in Fig.~\ref{muli7}. In this figure the two solid
purple lines correspond to the VMC (impulse and total) values, the green
squares (red circles) are extrapolated GFMC impulse (total) propagations
and the solid green (red) lines starting at $\tau$=0.1 MeV$^{-1}$ are the
final GFMC averages with dashed lines to denote the Monte Carlo error.
The GFMC propagation reduces the VMC matrix element slightly in both cases.
The GFMC impulse result is only 88\% of the experimental value,
but the MEC contributions raise the total to 97\%.

\subsection {Magnetic Dipole Transitions in $A$=6,7 Nuclei}

In Table~\ref{tb:m1} we present the different contributions to the matrix
elements for $M1$ transitions in $A$=6,7 nuclei.  As in the
case of magnetic moment calculations, we see that the most significant contributions
come from the pseudoscalar and vector pieces of the two-body current operators.

\begin{table}[bt]
\caption{Matrix elements of $M1$ transitions in $A$=6,7 nuclei.
Column headings are defined in Table~\ref{tb:mu}.}
\label{tb:m1}
\vspace{5mm}
\begin{ruledtabular}
\begin{tabular}{llcccccc}
$J^\pi_i\to J^\pi_f$& Method                           & IA       & \multicolumn{4}{c}{MEC}            & Total\\
      & &   & PS$+$V &MS &MD&$\Delta$ &\\
\colrule
$^6$Li$(0^+;1)\to^6$Li$ (1^+;0)$ &VMC     	     & ~3.683(14) & 0.307   & ~0.003 & ~0.010  &-0.053 & ~3.950(14)\\
$^6$Li$(0^+;1)\to^6$Li$ (1^+;0)$&GFMC 		     & ~3.587(16) & ~0.323  & ~0.002 & ~0.012  &-0.048 & ~3.876(14)\\
\\
$^7$Li$(\frac{1}{2}^-)\to^7$Li$(\frac{3}{2}^-)$&VMC  & ~2.743(17) &  0.396  & ~0.006 & -0.017  &-0.034 & ~3.162(22)\\
$^7$Li$(\frac{1}{2}^-)\to^7$Li$(\frac{3}{2}^-)$&GFMC & ~2.677(19) &  0.395  & ~0.011 & -0.017  & 0.072  & ~3.138(22)\\
\\
$^7$Be$(\frac{1}{2}^-)\to^7$Be$(\frac{3}{2}^-)$&VMC  & ~2.420(30) & ~0.390  & -0.005 & ~0.010  &-0.024 & ~2.791(36) \\
$^7$Be$(\frac{1}{2}^-)\to^7$Be$(\frac{3}{2}^-)$&GFMC & ~2.374(31) & ~0.394  & -0.010 & ~0.010  &-0.002 & ~2.766(36)\\
\end{tabular}
\end{ruledtabular}
\end{table}

\begin{figure}[ht!]
\centerline{\epsfig{file=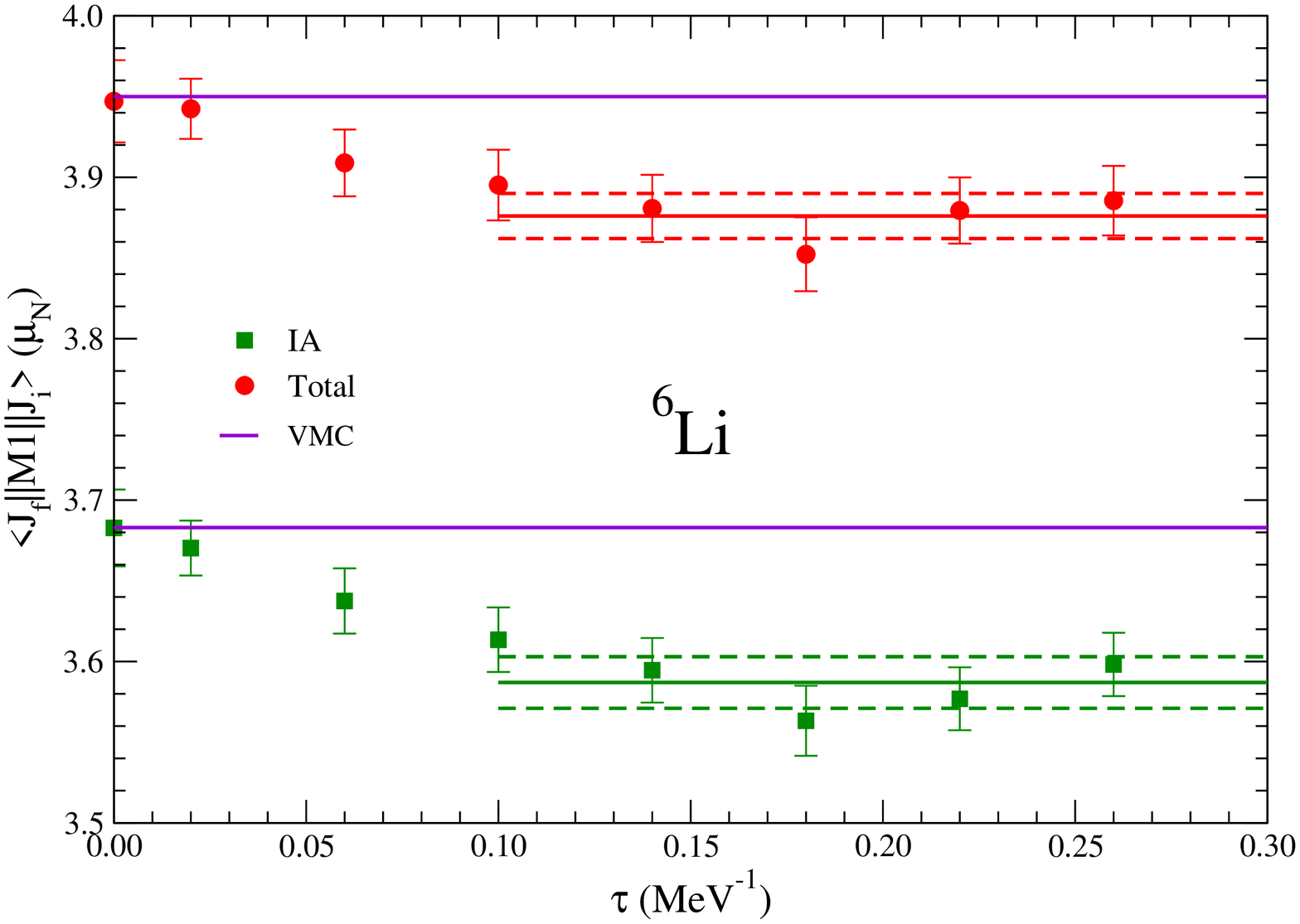,width=16.4cm}}
\caption{(Color online)
Extrapolated GFMC $M1$ matrix element for the $^6$Li$(0^+;1)\to^6$Li$(1^+;0)$ 
transition in impulse approximation (green squares) and with MEC (red circles).
VMC values (purple lines) and averaged GFMC values (lines with error bars)
are also shown.}
\label{m1li6}
\end{figure}

The first two rows of Table~\ref{tb:m1} show the various pieces of the $M1$ 
matrix element for $^6$Li$(0^+;1)\to ^6$Li$ (1^+;0)$ transition.  We note that
both VMC and GFMC IA results are boosted by 7--8\% by MEC.  The
corresponding decay widths for this transition are presented in
Table~\ref{tb:width}.  The total width we obtain from the VMC
calculation agrees very well with the experimental value,
whereas the total GFMC width is slightly outside of the
present experimental range.  Figure~\ref{m1li6} shows the matrix 
elements for this case as a function of $\tau$.  As in the previous figure, 
the two solid purple lines represent the VMC impulse and total estimates,
the green squares represent the GFMC propagated points for impulse and 
the red circles denote the total GFMC matrix elements.  The average GFMC 
results are shown as solid lines with error bars starting at 
$\tau$=0.1 MeV$^{-1}$.
All the GFMC points as well as the averages represent the extrapolated 
matrix elements.  We note that, also in the present case,
the GFMC propagation slightly decreases the VMC value for the
$^6$Li$(0^+;1)\to^6$Li$ (1^+;0)$ transition.  We see that the propagated 
points are quite stable with $\tau$, which we ran up to $0.3$ MeV$^{-1}$
in this case.

\begin{table}[bt]
\caption{Impulse approximation (IA) and total $M1$ transition widths in eV for $A$=6,7 nuclei.}
\label{tb:width}
\vspace{5mm}
\begin{ruledtabular}
\begin{tabular}{lcccccc}
$J^\pi_i\to J^\pi_f$&Mode                          &\multicolumn{2}{c}{VMC} &\multicolumn{2}{c}{GFMC}  &   Expt.  \\
                                               &                 &   IA   & Total  &   IA    & Total   &        \\
\colrule
$^6$Li$(0^+;1)\to^6$Li$ (1^+;0)$               & $M1$       &7.09(6)&8.15(6)&6.72(6)&7.85(6) &8.19(17) \\
$^7$Li$(\frac{1}{2}^-)\to^7$Li$(\frac{3}{2}^-)$& $M1$ ($10^{-3}$)&4.75(6)&6.31(9)& 4.52(6)&6.21(9) &6.30(31)  \\ 
$^7$Be$(\frac{1}{2}^-)\to^7$Be$(\frac{3}{2}^-)$& $M1$ ($10^{-3}$)&2.62(7)&3.49(9)&2.52(7)&3.42(9)  &3.43(45) \\
\end{tabular}
\end{ruledtabular}
\end{table}

Table \ref{tb:width} also shows two
$M1$ transitions in $A$=7 nuclei. The MEC corrections are 15--17$\%$ of
the $M1$ matrix elements obtained in both VMC and GFMC calculations.  The
model independent pieces (PS+V) are the largest contributions
and are also very similar for both $^7$Li and $^7$Be.
The decay widths for
$^7$Li$(\frac{1}{2}^-)\to^7$Li$(\frac{3}{2}^-)$ and
$^7$Be$(\frac{1}{2}^-)\to^7$Be$(\frac{3}{2}^-)$ $M1$ transitions are shown
in Table~\ref{tb:width}.  The total widths from both VMC and GFMC
match very well with the experimental decay widths.

\section {Conclusions}
\label{sec:conclusions}

In summary, we have reported results for the magnetic moments and
magnetic dipole transitions in nuclei with mass numbers $A \leq 7$.
The calculations have used essentially exact wave functions derived
from a realistic Hamiltonian that reproduces well the low-lying spectra
of these nuclei as well as of those in the mass range $A$=8--10.  Leading
terms in the nuclear electromagnetic current have been constructed to
satisfy current conservation with the two-nucleon potential, AV18, used in the Hamiltonian.
Additional contributions associated with the explicit presence of $\Delta$
isobar degrees of freedom have been accounted for by including, in an
approximate fashion, $\Delta$ components in the nuclear wave functions
with the transition-correlation-operator method.

Overall, the agreement between the calculated and experimental magnetic
moments and transition rates is quite satisfactory, particularly in the
isovector channel where differences between computed and experimental
amplitudes are $\simeq 1.5$\%.
On the other hand, in the isoscalar channel these differences 
seem to progressively become worse as the mass number increases; they
are about 1\% in deuteron, 4\% in $^3$He/$^3$H, and 10\% in $^7$Be/$^7$Li.
Of course, isoscalar transitions are suppressed both at the one- and
two-body levels: the IA current is proportional to the nucleon isoscalar
magnetic moment, which is five times smaller than the corresponding
isovector combination; leading two-body currents from pion-exchange and
$\Delta$ excitation have isovector character.
Two-body isoscalar contributions arise in the present study from short-range
mechanisms: the momentum-dependent components of the AV18, the $\rho\pi\gamma$
transition current, and renormalization corrections induced by $\Delta$ admixtures
in the wave functions, Eq.~(\ref{eq:rcc1}).

We conclude by noting that in a chiral effective-field-theory framework isoscalar
corrections are suppressed by $(Q/\Lambda_\chi)^2$ ($Q$ denotes a generic small
momentum and $\Lambda_\chi \simeq 1$ GeV is the chiral-symmetry-breaking scale) relative
to the leading-order (LO) IA current~\cite{Pastore08a,Pastore08b}. These
N$^2$LO corrections have been calculated in the deuteron and trinucleon isoscalar
magnetic moments, and are $\simeq 1$\% relative to LO but of opposite sign, so that
they increase the discrepancy between theory and experiment.  At N$^3$LO, or
$(Q/\Lambda_\chi)^3$, a number of isoscalar two-body currents originate from four-nucleon
contact interactions involving two gradients of the nucleon fields~\cite{Pastore08b}.
Their contributions to electromagnetic observables have yet to be calculated.   It will
be interesting to see whether these isoscalar currents as well as the corresponding
isovector ones up to N$^3$LO will improve the present picture.

\acknowledgments

The many-body calculations were performed on the parallel computers of the
Laboratory Computing Resource Center, Argonne National Laboratory.
This work is supported by the U. S. Department of Energy,
Office of Nuclear Physics, under contracts No. DE-AC02-06CH11357
(M.P., S.C.P., and R.B.W.) and No. DE-AC05-06OR23177 (R.S.)
and under SciDAC grant No. DE-FC02-07ER41457.


\begin{thebibliography}{99}
\bibitem{PPW07}
M. Pervin, S. C. Pieper, and R. B. Wiringa,
Phys. Rev. C {\bf 76}, 064319 (2007).

\bibitem{WSS95}
R. B. Wiringa, V. G. J. Stoks, and R. Schiavilla,
Phys. Rev. C {\bf 51}, 38 (1995).

\bibitem{PPWC01}
S. C. Pieper, V. R. Pandharipande, R. B. Wiringa, and J. Carlson,
Phys. Rev. C {\bf 64}, 014001 (2001).

\bibitem{PW01}
S. C. Pieper and R. B. Wiringa,
Annu. Rev. Nucl. Part. Sci. {\bf 51}, 53 (2001).

\bibitem{PVW02}
S. C. Pieper, K. Varga, and R. B. Wiringa,
Phys. Rev. C {\bf 66}, 044310 (2002).

\bibitem{PWC04}
S. C. Pieper, R. B. Wiringa, and J. Carlson,
Phys. Rev. C {\bf 70}, 054325 (2004).

\bibitem{P05}
S. C. Pieper,
Nucl. Phys. {\bf A751}, 516c (2005).

%
\bibitem{Car98} J.\ Carlson and R.\ Schiavilla,
                Rev.\ Mod.\ Phys.\ {\bf 70}, 743 (1998).

\bibitem{WS98}
R. B. Wiringa and R. Schiavilla,
Phys. Rev. Lett. {\bf 81}, 4317 (1998).

\bibitem{W91}
R. B. Wiringa,
Phys. Rev. C {\bf 43}, 1585 (1991).

\bibitem{PPCPW97}
B. S. Pudliner, V. R. Pandharipande, J. Carlson, S. C. Pieper, and
R. B. Wiringa,
Phys. Rev. C {\bf 56}, 1720 (1997).

\bibitem{MR2T2}
N. Metropolis, A. W. Rosenbluth, M. N. Rosenbluth, A. H. Teller, and E. Teller,
J. Chem. Phys. {\bf 21}, 1087 (1953).

\bibitem{C87}
J. Carlson,
Phys. Rev. C {\bf 36}, 2026 (1987).

\bibitem{C88}
J. Carlson,
Phys. Rev. C {\bf 38}, 1879 (1988).

\bibitem{WPCP00}
R. B. Wiringa, S. C. Pieper, J. Carlson, and V. R. Pandharipande,
Phys. Rev. C {\bf 62}, 014001 (2000).

\bibitem{Mar05} L.\ E.\ Marcucci, M.\ Viviani, R.\ Schiavilla, A.\ Kievsky,
                and S.\ Rosati,
                Phys.\ Rev.\ C {\bf 72}, 014001 (2005).
%
\bibitem{Ris89} D.\ O.\ Riska,
                Phys.\ Rep.\ {\bf 181}, 207 (1989).
%
\bibitem{Ris85} D.\ O.\ Riska,
                Phys.\ Scr.\ {\bf 31}, 107 (1985).
%
\bibitem{Sac48} R.G.\ Sachs, Phys.\ Rev.\ {\bf 74}, 433 (1948).
%
\bibitem{Mar98} L.\ E.\ Marcucci, D.\ O.\ Riska, and R.\ Schiavilla,
                Phys.\ Rev.\ C {\bf 58}, 3069 (1998).
%
\bibitem{Car90} J.\ Carlson, D.\ O.\ Riska, R.\ Schiavilla, and R.\ B.\ Wiringa,
                Phys.\ Rev.\ C {\bf 42}, 830 (1990).
%
\bibitem{Sch91} R.\ Schiavilla and D.\ O.\ Riska, 
                Phys.\ Rev.\ C {\bf 43}, 437 (1991).
%
\bibitem{Viv96} M.\ Viviani, R.\ Schiavilla, and A.\ Kievsky, 
                Phys.\ Rev.\ C {\bf 54}, 534 (1996).
%
\bibitem{Viv00} M.\ Viviani, A.\ Kievsky, L.\ E.\ Marcucci, S.\ Rosati, 
		and R.\ Schiavilla,
                Phys.\ Rev.\ C {\bf 61}, 064001 (2000).
%
\bibitem{Ber80} D.\ Berg {\it et al.},
                Phys.\ Rev.\ Lett.\ {\bf 44}, 706 (1980).
%
\bibitem{Che71} M.\ Chemtob and M.\ Rho,
                Nucl.\ Phys.\ {\bf A163}, 1 (1971).
%
\bibitem{WSA84} 
R. B. Wiringa, R. A. Smith, and T. L. Ainsworth,
Phys. Rev. C {\bf 29}, 1207 (1984).
%
\bibitem{Sch92} R.\ Schiavilla, R.\ B.\ Wiringa, V.\ R.\ Pandharipande, 
                and J.\ Carlson,
                Phys.\ Rev.\ C {\bf 45}, 2628 (1992).
%
\bibitem{Car86} C.\ E.\ Carlson, 
                Phys.\ Rev.\ D {\bf 34}, 2704 (1986).
%
\bibitem{Lin91} D.\ Lin and M.\ K.\ Liou, 
                Phys.\ Rev.\ C {\bf 43}, R930 (1991).
%
\bibitem{Eri88} T.\ E.\ O.\ Ericson and W.\ Weise, {\it Pions and Nuclei} 
                (Clarendon Press, Oxford, 1988).
%
\bibitem{exp567}
D. R. Tilley, C. M. Cheves, J. L. Godwin, G. M. Hale, H. M. Hofmann,
J. H. Kelley, C. G. Sheu, and H. R. Weller,
Nucl. Phys. {\bf A708}, 3 (2002).
%
\bibitem{WN08}
W. N\"ortersh\"auser {\it et al.}, arXiv:0809.2607.
%
\bibitem{PPCW95}
B. S. Pudliner, V. R. Pandharipande, J. Carlson, and R. B. Wiringa,
Phys. Rev. Lett. {\bf 74}, 4396 (1995).
%
\bibitem{FPPWSA96}
J. L. Forest, V. R. Pandharipande, S. C. Pieper, R. B. Wiringa, R. Schiavilla,
and A. Arriaga, 
Phys. Rev. C {\bf 54}, 646 (1996).
%
\bibitem{Pastore08a}
S.\ Pastore, R.\ Schiavilla, and J.L.\ Goity,
proceedings of the {\it Fourth Asia-Pacific Conference on Few-Body
Problems in Physics}, Depok, Indonesia, August 19--23, 2008, to be
published in Mod.\ Phys.\ Lett.\ A; arXiv:0809.2555.
%
\bibitem{Pastore08b}
S.\ Pastore, R.\ Schiavilla, and J.L.\ Goity,
in preparation.

\end{thebibliography}
\end{document}